\documentclass[12pt,epsf,twoside]{article} \input{epsf.sty}

\def\mypagenumber{1}
\def\mydate{May 2, 2000 }
\def\myend{\end{document}}

\def\Journal#1#2#3#4{{#1}{\bf #2} (#3) #4}
\def\CQG{\em Class.\ Quant.\ Grav.}
\def\NPB{{\em Nucl.\ Phys.} B}
\def\PLB{{\em Phys.\ Lett.} B}
\def\PRL{\em Phys.\ Rev.\ Lett. }

\def\PRD{{\em Phys.\ Rev.} D}
\def\AP{{\em Ann.\ Phys.\ (N.Y.)} }

\def\CMP{\em Comm.\ Math.\ Phys. }

\normalsize

\newcounter{sxn}

\newcounter{axn}

\date{}

\newdimen\mybaselineskip
\newcommand{\beeq}{\begin{equation}}
\newcommand{\eneq}{\end{equation}}
\newcommand{\be}{\begin{eqnarray}}
\newcommand{\ee}{\end{eqnarray}}
\newcommand{\bpic}{\begin{picture}}
\newcommand{\epic}{\end{picture}}

\def\dd{\partial}
\def\la{\raise.16ex\hbox{$\langle$} \, }
\def\ra{\, \raise.16ex\hbox{$\rangle$} }

\def\psibar{ \psi \kern-.65em\raise.6em\hbox{$-$} }
\def\mbar{ m \kern-.78em\raise.4em\hbox{$-$}\lower.4em\hbox{} }

\def\ep{\epsilon}

\def\n@space{\nulldelimiterspace=0pt \mathsurround=0pt }
\def\huge#1{{\hbox{$\left#1\vbox to 20.5pt{}\right.\n@space$}}}

\def\myskip{\noalign{\kern 8pt}}
\def\myeqspace{\noalign{\kern 10pt}}

\def\boxit#1{$\vcenter{\hrule\hbox{\vrule\kern3pt
    \vbox{\kern3pt\hbox{#1}\kern3pt}\kern3pt\vrule}\hrule}$}
\def\bigbox#1{$\vcenter{\hrule\hbox{\vrule\kern5pt
     \vbox{\kern5pt\hbox{#1}\kern5pt}\kern5pt\vrule}\hrule}$}

\def\ignore#1{{}}

\begin{document}

\bibliographystyle{unsrt}
\footskip 1.0cm

\thispagestyle{empty}
\setcounter{page}{\mypagenumber}

%{\baselineskip=10pt \parindent=0pt \small
%\mydate  
%}                
             
\begin{flushright}{\mydate ~ (.)
\\  UMN-TH-1841/00  , TPI-MINN-00/9-T \\
,OUTP-00-02-P\\
}

\end{flushright}

\vspace{2.5cm}
\begin{center}
{\LARGE \bf {Gravitating Instantons In 3 Dimensions
}}\\ 
%\vspace {1.0cm}
%{\LARGE  in  3 Dimensions }}\\
\vspace{2cm}
{\large Andrew Ferstl$^a$, Bayram Tekin$^{b,}$\footnote{e-mail:~
tekin@thphys.ox.ac.uk}, Victor Weir$^a$
}\\
\vspace{.5cm}
$^a${\it School of Physics and Astronomy, University of Minnesota,
         Minneapolis, MN 55455}\\
\vspace{0.5cm}
$^b${\it Theoretical Physics, University of Oxford, 1 Keble Road, Oxford,
OX1 3NP, UK}\\  
\end{center}

\vspace*{2.5cm}
%\baselinestretch{2.0}

%\normalsize

\begin{abstract}
\baselineskip=18pt
We study the Einstein-Chern-Simons gravity coupled to Yang-Mills-Higgs 
theory in three dimensional Euclidean space with cosmological constant. 
The classical equations reduce to  Bogomol'nyi type first order equations in curved space.
There are BPS type gauge theory instanton (monopole)  solutions of finite 
action in a gravitational instanton which itself has a finite action. We also discuss 
gauge theory instantons in the vacuum (zero action) AdS space.
  In addition we point out to some exact solutions which are singular. 
\end{abstract}
\vfill



%\end{titlepage}
 
\newpage

%\setcounter{page}{1}

%\textheight=20cm
%\headsep=0.75cm
%\vsize=20cm

%%%%%%%%%%%%%%%%%%%%%%%%%%%%%%%%%%%%%%%%%%%%%%%%%%%%%%%%%%%%%%%%%
\normalsize
\baselineskip=22pt plus 1pt minus 1pt
\parindent=25pt
%\vspace*{5mm}

\section{Introduction}

In recent years there has been a growing interest in the study 
of three dimensional gravity \cite{Deser1}. 
For example the BTZ \cite{BTZ} solution in three dimensions proved to be an
extremely useful toy model to understand microscopic degrees of freedom
of a black hole. Quite recently \cite{Teitelboim} a black hole with all
three Abelian hairs (charge, angular momentum and mass ) was shown to
exist.

On the other hand, since the surprising numerical evidence of BK
\cite{BK} on the existence of particle-like solutions (with hair) 
in four dimensional Einstein-Yang-Mills theory, there has been a 
lively research in the theories of gravity coupled to non-Abelian 
gauge theories. A nice account of developments and references in the
subject 
is summarized in the review article \cite{Volkov}. More recently 
new monopole and dyon solutions were found in \cite{Winstanley}.

In this paper our intention is to study three dimensional Euclidean 
gravity coupled to Yang-Mills-Higgs theory. Deser's \cite{Deser} earlier
work in the study of Einstein-Yang-Mills theory (no Higgs) shows that
there are no static solutions in this theory. In this paper we add a
Higgs field in the adjoint representation of the group $SO(3)$ which is
spontaneously broken down to $U(1)$ and we study the effect of gravity
on the 't Hooft-Polyakov instantons in the BPS limit (the limit of 
vanishing self-interaction for the Higgs field). We show that
one obtains the Bogomol'nyi type first order equations for the Higgs and the
gauge field as in the flat space limit. There are exact solutions to the
equations of motion which do not have  flat space analogs
(limits)  but these solutions are not of finite action. We find the
numerical solutions of finite action which reduce to
the BPS instantons in flat space. The most interesting thing that 
we found is that gauge theory instantons exist in  gravitational instantons.
Gravitational instantons are not disturbed by the the gauge theory instantons.

The reader might wonder if a spontaneously
broken, $ SO(3)$ down to $U(1)$, gauge theory is expected to be any
different from the Einstein-Maxwell theory in which a BTZ solution was found. 
In the four dimensional context it \cite{Greene} was shown that, on the
contrary to the initial expectation, the spontaneously broken gauge theory
allowed  black holes with ``non-trivial'' hair which do not exist in
Einstein-Maxwell theory.
So, in principle, we do not have a strong reason to believe that the
spontaneously broken theory, in three dimensions, will only have a BTZ type
solution with only three types of hair. Although this question along with
the question of non-Abelian hair of a BTZ black hole are extremely
interesting we do not attempt these in this paper. Our immediate 
interest, as stated in the previous paragraph, is to explore what 
happens to the
gauge theory instantons if gravity ( with cosmological constant) is turned
on. And more specifically we will explore what happens to the gauge theory
instantons in the case that the space-time is a gravitational instanton.  
In section 2 we present the model in which we will search for 
an answer to this problem.  In section 3 we will begin to solve the 
equations of motion, discuss some of the remaining symmetries, and show that 
the action does not depend on a specific choice of the 
coordinates.  In section 4 we will find numerical solutions.  
In section 5 we will present an exact,
 singular solution to the equations of motion and finally in Section 6
we will make some concluding remarks.
 
\section{The Model}

We will work in the Euclidean space and in the first order formalism of
gravity in terms of the dreibein and the spin connection.
Ach\'{u}carro-Townsend \cite{Townsend} and Witten \cite{Witten} showed
that three dimensional Einstein-Hilbert action with zero cosmological
constant is equivalent to Chern-Simons theory with the gauge group
$ISO(3)$. In the theories with non-zero cosmological constant one can
simply generalize this to $SO(4)$ or $SO(3,1)$ depending on the sign
of the cosmological constant. Witten also realized that 
depending on the choice of the quadratic Killing-form one obtains two 
classically equivalent actions for gravity.   

The standard action is the following~\footnote{In our conventions the integrand of the Euclidean
 path integral is $e^{+S}$.}

\be
S_G = -{1\over{16\pi G}} \int_{\cal{M}} d^3x \epsilon^{ijk}\Bigg\{2\,
e^a\,_i \,\dd_j\,\omega^a\,_k + \epsilon^{abc}\,e^a\,_i
\,\omega^b\,_j\,\omega^c\,_k +
{\lambda\over 3}\epsilon^{abc}\,e^a\,_i\,e^b\,_j\,e^c\,_k \Bigg\}
\ee
This action is real in the Euclidean space and is equivalent to
Einstein-Hilbert theory if the dreibein is invertible.  Another action,
 which in Minkowski space
 is equivalent to the Einstein-Hilbert action, 
often coined as "exotic action" is:  
\be
S_{CS} = 
&&-{ik\over{8\pi G}} \int_{\cal{M}}d^3x \epsilon^{ijk}\Bigg\{\,
\omega^a\,_i \,\dd_j\,\omega^a\,_k + {1\over
3}\epsilon^{abc}\,\omega^a\,_i
\,\omega^b\,_j\,\omega^c\,_k +
\lambda (e^a\,_i\dd_j\,e^a\,_k \nonumber \\
&&+\epsilon^{abc}\,\omega^a\,_i\,e^b\,_j\,e^c\,_k) \Bigg\}
\ee
Assuming that the fields are real, which
we do for our analysis, this action is completely imaginary
because it is Wick rotated from Minkowski
space. Both of these actions are considered in the path integral
formulation of gravity. These actions are important geometric invariants
of three manifolds; namely, they are the ''volume" and the ''Chern-Simons"
invariants respectively. In this paper we consider both of them together.  
So including the Higgs and the Yang-Mills terms our full action becomes
   
\be
S= S_G + S_{CS} + S_{YM} + S_H
\ee
where the Yang-Mills and the Higgs actions are
\be
S_{YM}= - {1\over{4e^2}}\int d^3x
\sqrt{|g|} g^{ij}g^{kl}F^a_{ik}F^a_{jl}
\ee
\be
S_H= -{1\over e^2} \int d^3x \sqrt{|g|} \{ {1\over 2} g^{ij}D_i h^a D_j
h^a +{\nu\over
6!}(h^2-h_0^2)^3 \}
\ee

The Higgs field 
is in the adjoint representation of $SO(3)$ and the covariant
derivative is $ D_i h^a = \dd_i h^a + \epsilon^{abc} A_i^b h^c$.  Hence,
$F^a_{ij}$ has no gauge coupling in it.

Let us denote the dimensions of the fields and the parameters in the
theory.
\be
&&[e^2] = M , \hskip  1cm [G] = M^{-1} , \hskip 1 cm  [\lambda] = M^2, 
\hskip 1cm [k]= M^{-1} \nonumber  \\
&&[e^a\,_j] = M^0 , \hskip  1cm [h] = M , \hskip 1 cm  [\omega^a\,_j] = M,
\hskip 1 cm  [\nu] = M^{-2}
\ee

We will exclusively work in the BPS limit where $\nu= 0$. 
The indices $(a,b,c)$ denote the tangent space and
$(i,j,k)$ denote the manifold coordinates. The metrics , $\eta_{ab}$
and $g_{ij}$ have Euclidean signature. $\lambda < 0 $ corresponds to the
de-Sitter and $\lambda > 0$ to the anti-de-Sitter space. The ``dual"
Riemann tensor can be defined to be $R^a\,_{kj} = \dd_k \omega^a\,_j
-\dd_j
\omega^a\,_k +\epsilon^a\,_{bc}\omega^b\,_k \omega^c\,_j $. The relation
between the
Ricci tensor and the dual Riemann tensor is 
$R_{ij} = e^a_i E^k_b \epsilon_{abc}R^c\,_{jk}$, where
$ E^k_b $ is the inverse of the dreibein. In the absence of gauge
fields Einsteins equations $R_{ij}= -2\lambda g_{ij}$ imply the scalar
curvature to be $R = -6 \lambda$. 

We employ the well known spherically symmetric ansatz for all the fields
in the theory.
\be
&&e^a\,_j(\vec{x})= {G\over r} \left[ -\epsilon^a\,_{jk}\,
\hat{x}^k\,\phi_1 + \delta^a\,_j\,\phi_2 +(r A-\phi_2)\,\hat{x}^a
\hat{x}_j \right]\\          
&&w^a\,_j(\vec{x})= {1\over r} \left[ \epsilon^a\,_{jk}\,
\hat{x}^k\,(1-\psi_1)+ \delta^a\,_j\psi_2 +(r B-\psi_2)\hat{x}^a
\hat{x}_j\right]\\
&&A^{a}_{j}(\vec{x}) = {1\over r} \left[ \epsilon^a\,_{jk}\,
\hat{x}^k\,(1-\varphi_1)+ \delta^a\,_j\varphi_2 +(r D-\varphi_2)\hat{x}^a
\hat{x}_j\right]\\
&&h^a(\vec{x})= \hat{x}^a h(r)
\ee

The functions $A$, $B$, $\phi_{\alpha}$ ,$D$, 
$\varphi_{\alpha}$ and $\psi_{\alpha}$ depend on $r$ only. The
meaning of $r$ should be clear from $ r^2= \eta_{ij}x^i x^j$ and we
define $\hat{x}^j = x^j/r$.
We have chosen the dreibein to be dimensionless and the first term of the
dreibein is chosen in a way which will yield more transparent equations.

The metric on the manifold can be recovered from the dreibein through
the relation 
$ g_{ij} = \eta_{ab} e^a\,_i e^b\,_j$ which yields; 
\be
g_{ij} = {G^2\over r^2}\Bigg\{ (\phi_1^2 + \phi_2^2)
(\delta_{ij} -\hat x_i \hat x_j)
         + r^2 A^2 \hat x_i \hat x_j\Bigg\}   
\ee
The flat space limit ($ g_{ij} = \delta_{ij}$) corresponds to  
$G^2 (\phi_1 ^2 + \phi_2^2) = r^2$ and $A(r)G= 1$. 

The dual Riemann tensor and the non-Abelian field strength tensor can be
obtained from a tedious but straightforward computation. Clearly both of
them are of the same form.

\be
R^a\,_{ij} &=&
 {1\over r^2}
  \ep_{ij b}\,  \hat x^a \hat x^b\, (\psi_1^2 + \psi_2^2 -1)
+ {1\over r} (\ep^a\,_{ij} -  \ep_{ij b}  \hat x^a \hat x^b )
(\psi_1' + B\psi_2) \cr
\noalign{\kern 10pt}
&& \hskip 4cm + (\delta^a\,_j \hat x_i - \delta^a\,_i \hat x_j)
{1\over r}(\psi_2' - B\psi_1)
\label{Riemann tensor}
\ee

\be
F^a\,_{ij} &=&
 {1\over r^2}
  \ep_{ij b}\,  \hat x^a \hat x^b\, (\varphi_1^2 + \varphi_2^2 -1)
+ {1\over r} (\ep^a\,_{ij} -  \ep_{ij b}  \hat x^a \hat x^b )
(\varphi_1' + D\varphi_2) \cr
\noalign{\kern 10pt}
&& \hskip 4cm + (\delta^a\,_j \hat x_i - \delta^a\,_i \hat x_j)
{1\over r}(\varphi_2' - D\varphi_1)
\ee
Before we write down the reduced form of the action let us denote the
determinant of the metric
\be
\det e = \sqrt{|g|}= {G^3\over r^2}|A|(\phi_1^2 +\phi_2^2)
\ee

The actions reduce to the following one dimensional forms
(here the repeated greek indices take values of $(1,2)$ and a
summation is implied).
 The Einstein-Hilbert action is

\be
S_G = - \int_0^{\infty} dr \, \Bigg\{
 \psi_{\alpha}' \epsilon_{\alpha \beta}\phi_{\beta} + B\psi_{\alpha}
 \phi_{\alpha} +
 {A\over 2}(\psi_{\alpha} \psi_{\alpha} +\lambda G^2 
\phi_{\alpha} \phi_{\alpha} -1) \Bigg \}
\ee
The Chern-Simons action is
\be
S_{CS} =  &&- i {k\over G}\int_0^{\infty} dr \, \Bigg\{
 \psi_{\alpha}' \epsilon_{\alpha \beta}\psi_{\beta} +\psi_2' + 
B(\psi_{\alpha} \psi_{\alpha} +\lambda G^2
\phi_{\alpha}\phi_{\alpha} -1)\nonumber \\
&&+\lambda G^2(\phi_{\alpha}'\epsilon_{\alpha \beta} \phi_{\beta} +2 
A \phi_{\alpha}\psi_{\alpha} )
 \Bigg\}
\ee

\be 
S_{YM}=- {2\pi\over{e^2G}}\int_0^{\infty} dr{1\over{|A|
\phi_{\delta}\phi_{\delta}}}\Bigg\{
A^2(\varphi_{\alpha}\varphi_{\alpha} -1)^2 
+2\phi_{\gamma}\phi_{\gamma}( \varphi_{\beta}'\varphi_{\beta}' \nonumber \\
+ 2D\epsilon_{\alpha \beta}\varphi_{\alpha}'\varphi_{\beta}
+D^2
\varphi_{\alpha}\varphi_{\alpha})\Bigg \}
\ee
 
In the BPS limit ($\nu=0$) and the broken phase ($h_0 \neq 0$) the Higgs
term is

\be
S_H=- {2\pi G\over e^2}\int_0^{\infty} dr\Bigg\{ {1\over{|A|}} 
\phi_{\alpha}\phi_{\alpha} h'^2
+ 2h^2
|A|\varphi_{\alpha}\varphi_{\alpha} \Bigg \}
\ee

From here we will assume that A(r) is positive so that we may drop the 
absolute value sign.  We will see that this requirement is satisfied
in our solution.

We are interested both in the singular and the  non-singular solutions.    
For the case of finite action and non-singular solutions
the boundary conditions for the gauge and Higgs sector follow as

\be 
&&\varphi_1(0)= 1 ,\hskip 0.5 cm \varphi_2(0)= 0 \hskip 1 cm
\varphi_1(\infty)= \varphi_2(\infty)= 0 \\
&& h(0)= 0 \hskip 1 cm h(\infty)= h_0 \hskip 1cm D(\infty)= 0
\ee

The equations of motion of the full theory are 

\be
\delta B:\hskip 2 cm  \psi_{\alpha} \phi_{\alpha} + i{k\over G} 
(\psi_{\alpha} \psi_{\alpha}  +
\lambda G^2\phi_{\alpha} \phi_{\alpha} -1) = 0
\ee
\be
\delta \psi :\hskip 2cm  \epsilon_{\alpha \beta}\phi_{\beta}' 
-B\phi_{\alpha} -A\psi_{\alpha} 
+i{k\over G}(2\epsilon_{\alpha \beta}\psi_{\alpha}' 
+2B\psi_{\beta} +2A\lambda G^2\phi_{\beta}) = 0
\ee
\be
\delta D : \hskip 2cm \epsilon_{\alpha \beta}\varphi_{\alpha}'
\varphi_{\beta}  +D\varphi_{\alpha} \varphi_{\alpha} = 0
\ee

\be
\delta h :\hskip 1cm \Bigg\{{\phi_{\alpha}\phi_{\alpha} h'\over A} \Bigg\}'
  -2 h A\varphi_{\alpha} \varphi_{\alpha} = 0
\ee

\be
\delta \phi :\hskip 0.5 cm &&\epsilon_{\alpha \beta}\psi_{\alpha}' 
+B\psi_{\beta} +A\lambda 
G^2\phi_{\beta}  
-i{k\over G}\lambda G^2(2\epsilon_{\alpha \beta}\phi_{\alpha}' 
-2B\phi_{\beta} -2A\psi_{\beta})\\ \nonumber
&&- {4\pi\over {e^2G}}{\phi_{\beta} A\over{(\phi_{\gamma}\phi_{\gamma})^2}}
(\varphi_{\alpha}\varphi_{\alpha} -1)^2+ {4\pi G\over e^2}
{\phi_{\beta} h'^2\over A}
=0
\ee

\be 
\delta \varphi :\hskip 0.5cm 
\Bigg\{ {{\varphi_{\alpha}'+D\epsilon_{\alpha \beta}\varphi_{\beta}}\over A} 
\Bigg\}' 
-{{\varphi_{\alpha}  A}\over {\phi_{\gamma}\phi_{\gamma}}}(\varphi_{\beta} 
\varphi_{\beta}-1) 
 -{D\over A}(D \varphi_{\alpha}+\epsilon_{\beta \alpha}\varphi_{\beta}') 
-G^2h^2A\varphi_{\alpha} =0 
\ee

\be
\delta A :\hskip 1 cm
&&\psi_{\alpha} \psi_{\alpha}  +\lambda G^2\phi_{\alpha} \phi_{\alpha} 
-1 +4ik\lambda G\phi_{\alpha} \psi_{\alpha} 
+{{8\pi G}\over e^2}h^2\varphi_{\alpha} \varphi_{\alpha}
+{{4\pi}\over {G e^2}}{(\varphi_{\alpha} \varphi_{\alpha}-1)^2\over{
\phi_{\gamma} \phi_{\gamma}}}\nonumber \\
&&-{{4\pi}\over {G e^2}}{1\over A^2}
\Bigg\{2\varphi_{\alpha}' \varphi_{\alpha}' +
4D \varphi_{\alpha}' \epsilon_{\alpha \beta} \varphi_{\beta} 
+ 2D^2\varphi_{\alpha} \varphi_{\alpha} 
+G^2h'^2\phi_{\alpha} \phi_{\alpha} \Bigg\}      
=0
\ee

In general, because of the Chern-Simons term, solutions to
the equations of motion will be complex. Complex dreibein and 
spin connection, however,  will
change the geometry drastically. For example, the notion of a positive
definite metric will be lost.
Hence, we restrict ourselves to the real solutions
only. This means that the Chern-Simons term decouples from the rest.
The equations of motion for the Chern-Simons gravity
are exactly the equations one gets for Einstein-Hilbert gravity without
the matter fields. This fact is no secret because we know that at the
classical level Einstein-Hilbert theory is equivalent to Chern-Simons
theory of gravity. In this way we have obtained a nice system where we
can try to analyze the effect of gravity on three
dimensional 't Hooft-Polyakov Instantons. It is clear that gravity
itself is not disturbed by the instantons because of the Chern-Simons
term. A similar situation arises in four dimensional Euclidean 
Einstein-Yang-Mills theory \cite{Duff}.  Charap and Duff showed that
in the $4D$ theory gauge theory instantons, having 
a vanishing energy momentum tensor, do not disturb the geometry. But the 
effect of gravity on the instantons is quite drastic.    
Using these facts we take on the job of obtaining solutions 
to the equations in the next section.

\section{Solutions of the Equations of Motion}

As already stated, we are only looking for real solutions, hence 
the Chern-Simons term yields the following equations:
\be
\label{eqn1}
\epsilon_{\alpha \beta}\phi_{\beta}' - A\psi_{\alpha} - B\phi_{\alpha}  = 0 \\
\label{eqn2}
\epsilon_{\alpha \beta}\psi_{\beta}' - B \psi_{\alpha} -
\lambda G^2A\phi_{\alpha} = 0 \\
\label{eqn3}
\psi_{\alpha} \psi_{\alpha} + \lambda G^2\phi_{\alpha}\phi_{\alpha} - 1 = 0 \\
\phi_{\alpha} \psi_{\alpha} = 0 
\label{orthogonal}
\ee
The general solutions of these equations, compatible with the regularity
conditions at the origin, were given in \cite{tekin}
\be
\psi_1 = {1\over\sqrt{1+\lambda G^2 f^2(r)}}\cos\Omega(r)&& \hskip 1 cm
\psi_2 = {1\over\sqrt{1+\lambda G^2 f^2(r)}}\sin\Omega(r) \\
\phi_1 = f(r){1\over\sqrt{1+\lambda G^2 f^2(r)}} \sin\Omega(r)&& \hskip
0.2cm
\phi_2 = - f(r) {1\over\sqrt{1+\lambda G^2 f^2(r)}}\cos\Omega(r)\\
A= - {f(r)'\over {1+\lambda G^2 f^2(r)}}&&\hskip 2.5 cm  B= \Omega'(r)
\ee
$f(r)$ and $\Omega(r)$  are arbitrary functions at this point. At the level
of the classical equations of motions one can pick any functions.
When we compute
the actions for the gravity sector we will see that their boundary values, 
namely $f(0)$, $f(\infty)$, $\Omega(0)$, $\Omega(\infty)$ are of extreme 
importance for the quantum theory.

We have postponed the issue of gauge fixing up until now. 
We need to see ``how arbitrary''  $f(r)$ and $\Omega(r)$ are and
whether we can gauge-fix any of them. A look at the equations of motion
will reveal that there is a remaining $U(1)$ symmetry 
which is not broken by the instanton ansatz. 
Under this symmetry the fields transform in the following way
\be
\tilde{\phi_1} &=& \phi_1 \cos\theta(r) +\phi_2 \sin\theta(r) \nonumber \\
\tilde{\phi_2} &=& -\phi_1 \sin\theta(r) +\phi_2 \cos\theta(r) \nonumber \\
\tilde{\psi_1} &=& \psi_1 \cos\theta(r) +\psi_2 \sin\theta(r) \nonumber \\
\tilde{\psi_2} &=& -\psi_1 \sin\theta(r) +\psi_2 \cos\theta(r) \nonumber \\
\tilde{B} &=& B - \theta'(r) \hskip 1cm \tilde{A} = A
\ee
So $f(r)$ is intact under this symmetry but $\Omega(r)$ is transformed.~\footnote{
All the actions except the Chern-Simons term is invariant under these transformations.
Chern-Simons term transforms like $\delta S_{CS}= -{ik\over G} (\gamma(\infty) -\gamma(0) )$.
As a compact subgroup of $SO(3)$ the remaining $U(1)$ can be parameterized 
$g(\vec{x})= e^{i\gamma(r) x^i \sigma^i}$. Where $\sigma^i$ are the Pauli matrices. If one is working
in a compact space, like $S^3$, for the gauge invariance of the path integral $k/G$ will be quantized.
This is because $ (\gamma(\infty) -\gamma(0) )$ is the winding number of $g(\vec{x})$. On the other
hand in an open ball like the one we deal with in this paper the Chern-Simons 
 coefficient is not quantized.
In this case  $ (\gamma(\infty) -\gamma(0) )$ becomes a collective coordinate which should be
summed over in the path integral.  See the discussion in \cite{Kamran}.}
$\theta(r)$ is the gauge parameter.
Choosing $\theta(r) = \Omega(r) $ one can work with the  following gauge 
equivalent fields
\be
&& \tilde{\psi_1} =  {1\over\sqrt{1+\lambda G^2 f^2(r)}}\, , \hskip 3cm 
\tilde{\psi_2} =0 \nonumber \\
&& \tilde{\phi_2} =  -f(r) {1\over\sqrt{1+\lambda G^2 f^2(r)}}\, , 
 \hskip 2cm \tilde{\phi_1} =0 \nonumber \\
&& A= - {f(r)'\over {1+\lambda G^2 f^2(r)}} \, ,\hskip 3.2 cm  \tilde{B}= 0
\ee

The line element in the polar coordinates takes the following form.
\be
(d s)^2 = {G^2\over{1+\lambda G^2 f^2}}\Bigg\{ f^2 d\Omega_2 +
{1\over{1+\lambda G^2 f^2}} (d f)^2 \Bigg\}
\ee

Once again we should emphasize that the spaces that are described by 
this metric are constant curvature spaces which satisfy 
$R^a_{ij}= -\lambda \epsilon^a_{bc}e^b_i e^c_j$ and  $R = - 6 \lambda$.
These are local properties of the space times. In the quantum theory 
global properties of the space time are needed. Below we will show that
the above metric describes many global-y inequivalent space-times depending
on the choice of the boundary values of $f(r)$. In terms of the    
gauge theory language: the space of $f(r)$ functions 
have a non-trivial topology. Before we start the discussion
of the actions in the gravity sector let us find the simplified equations
of the gauge sector.

Using the solutions of the Chern-Simons
equations of motion to simplify the equations for the
Higgs and Yang-Mills fields, we can see that the resulting relations
are still somewhat complicated.
To see the solution more clearly one can make certain choices of gauges.
For example a look at the action will reveal that the unbroken $U(1)$
acts in a way that  keeps the following complex function invariant
\be
\eta(r)= (\varphi_1 +i\varphi_2)e^{-i \int^r  D(r')dr'}
\ee
If the  Yang-Mills action is written in terms of $\eta(r)$ obviously none
of the functions $( \varphi_1, \varphi_2 , D)$ will appear in the action.
So we can choose a gauge ( the singular gauge ) in which $\varphi_2 = D
=0$. Denoting $\varphi_1 = \varphi$, the remaining {\it independent}
equations read as 
   
\be
h' = -{A\over {G (\phi_{\alpha}\phi_{\alpha})}}(\varphi^2 -1)
\label{hprime}
\ee

\be
h = -{1\over G A} {\varphi'\over{\varphi}} 
\label{varphiprime} 
\ee

 These are the Bogomol'nyi type first order equations 
for the gravitating instanton. These equations reduce to the well known
exactly solvable equations in the flat space limit. 
Writing 
the above equations explicitly in terms of the solutions of the gravity part
one obtains 
\be
h' = -{|f'(r)|\over {G f(r)^2}} (\varphi^2 -1)
\ee

\be
h= -{ \Big( 1+\lambda G^2 f(r)^2 \Big )\over {G|f'(r)|}}
{\varphi'\over{\varphi}} 
\ee

The coordinate function  $f(r)$ explicitly enters in the equations
so one should make sure that the existence of the solutions does not 
depend on the local coordinates. But the global properties of the space time
will be important as it should be expected. For generic $f(r)$ there
are exact solutions which will be depicted in the next section. But these are
all infinite action. Finite actions solutions will be found numerically. 
But first one needs to make a choice of $f(r)$

We write the gauge sector action in the following form
\be
 S_{YM} + S_H =&& - {4\pi\over{e^2G}}\int dr\Bigg\{ {1\over A} \Big( \varphi'+
G A h\varphi \Big)^2 - 2 G \varphi' \varphi h \nonumber \\
&& + {G^2 \over{2 A}}
\phi_{\alpha}\phi_{\alpha} \Big (h'+ {A\over {G
\phi_{\gamma}\phi_{\gamma}}(\varphi^2 -1)} \Big )^2 
- G h' (\varphi^2 -1) \Bigg \}
\label{lambdaeqn}
\ee
If the equations of motion are satisfied the integrand becomes a full
derivative and after integration one obtains
\be
S_{Instanton} = - {4\pi h(\infty)\over e^2}
\label{exactaction}
\ee
The result does not  explicitly dependent on the cosmological constant. 
It, however, can be seen from the numerical solutions that $h(\infty)$
depends on the cosmological constant.

Einstein-Hilbert action can be computed to be
\be
S_G= -\lambda G^2 \int_0^\infty dr {{|f'(r)| f(r)^2}
\over{(1+\lambda G^2 f(r)^2)^2}}
\ee
Observe that the integrand is a total derivative which can integrated to give
\be
S_G = {1\over{2 G \sqrt{\lambda}}} \Bigg\{ 
{ -G \sqrt{\lambda}|f(\infty)|\over{ 1 + G^2 \lambda f^2(\infty)}} +
{G \sqrt{\lambda}|f(0)|\over{ 1 + G^2 \lambda f^2(0)}}+
  \arctan{[G\sqrt{\lambda}|f(\infty)|]} 
- \arctan{[G\sqrt{\lambda}|f(0)|]}\Bigg\}   
\ee

It is clear that homotopically inequivalent $f(r)$'s characterize
different spaces. In what follows we will work on two different
spaces. The first one is given by $f(r)= - r/G$. The gravitational
action reads as~\footnote{ The Chern-Simons action will be 
 $S_{CS}= -{ik\over G} (\Omega(\infty) -\Omega(0) )$ which is exactly equal to the
gauge non-invariant part.}     
\be
S_G= - {\pi\over{4 \sqrt{\lambda}G}} 
\label{exactaction2}
\ee

This solution is a gravitational instanton ( of course it is not self-dual) and it is not
a gauge copy of the AdS space which has a zero action. The trivial vacuum AdS solution is
\be
G f(r)= -r/(1 - {\lambda \over 4}r^2)
\ee
which has a zero action.

\section{Numerical Computations}

For the gravitational instanton solution where $Gf(r)= -r$ the curved space BPS 
equations become
 \be
h'(r) = -\frac{1}{r^2} (\varphi^2(r) -1)
\ee

\be
\varphi'(r) = -{1\over( 1 + \lambda r^2)} h(r) \varphi(r)
\ee

and the line element becomes 
\be
(d s)^2 = {1\over{1+ {\lambda \over 4}r^2}}\Bigg\{ r^2 d\Omega_2 +
 \frac{(d r)^2}{1 + \lambda r^2} \Bigg\}
\ee

\begin{figure}
\begin{center}  
\leavevmode
\epsfsize= 8.0in \epsfbox{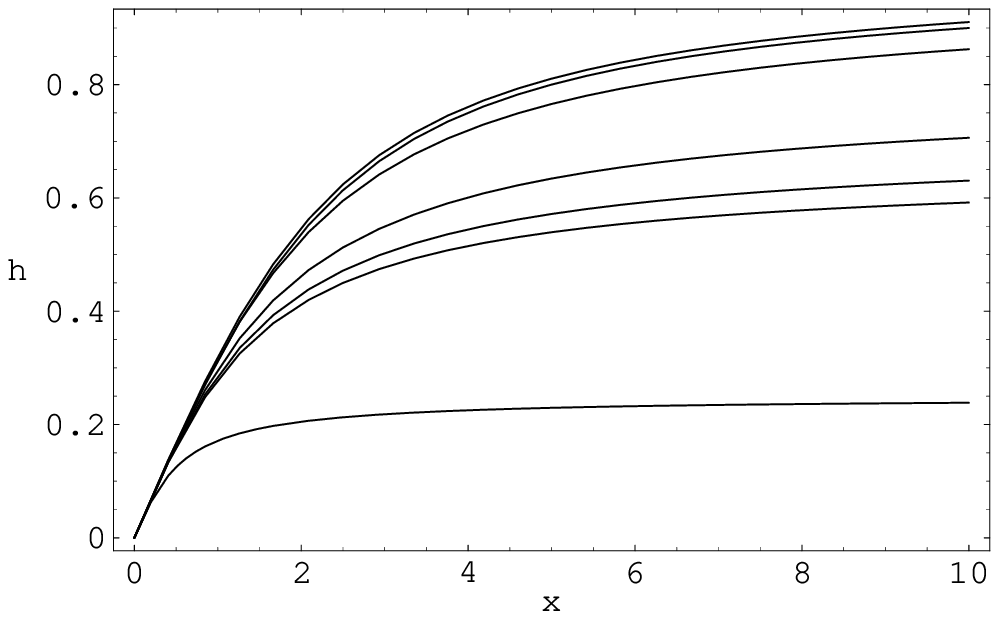}
\end{center}
\caption{ Higgs field $h(x)$ is shown for various values of $\lambda$
in the choice $G f = - r$.
$h(\infty)$ approaches to zero (vacuum solution) if $\lambda$ is
increased. The above
values, starting from the top correspond to $\lambda= 0$, $4. 10^{-3}$
,$\lambda=0.1$, $\lambda=0.5$, $\lambda= 0.8$, $\lambda= 1$ and
$\lambda=10$  in units of $h_0^2$.}
\label{fig1}
\end{figure}

\begin{figure}
\begin{center}
\leavevmode
\epsfsize= 8.0in \epsfbox{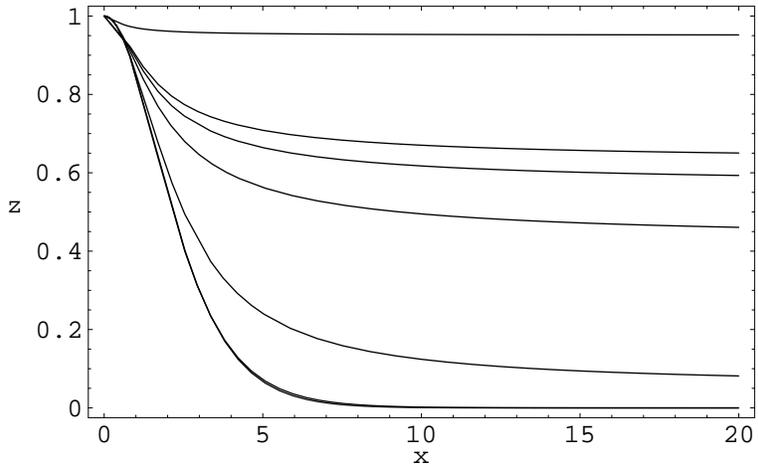}
\end{center}
\caption{ Non-zero component of the gauge field  $z(x)$ is shown for
various values of $\lambda$ in the choice $G f = -r$.
$z(\infty)$ approaches to 1 (vacuum solution) if $\lambda$ is increased.
The above
values, starting from the bottom correspond to $\lambda= 0, 4. 10^{-3}$
,$\lambda=0.1$, $\lambda=0.5$, $\lambda= 0.8$, $\lambda= 1$ and
$\lambda=10$ in units of $h_0^2$.}
\label{fig2}
\end{figure}

In the flat space limit $(\lambda =0)$ one has the well-known BPS solution
\be
\varphi(r)= {h_0 r\over\sinh(h_0 r)} ; \hskip 2cm h(r) = -{1\over r} +h_0
\coth(h_0 r)   
\label{BPS}
\ee
where $h(\infty) = h_0 $.  
For non-zero $\lambda$ the solutions can be obtained numerically and they
are plotted in figure \ref{fig1} and figure \ref{fig2}. 
 For any positive value of
$\lambda$ there is a solution. Non-zero $\lambda$ solutions take
values between the BPS instanton solution, (\ref{BPS}), and the trivial vacuum 
solution $( h(r) = 0, \varphi(r) = 1)$.  
For very large values of $\lambda$ the solution approaches the
trivial vacuum solution.  For negative values of $\lambda$ (i.e. the 
de Sitter case with our conventions) the existence of the horizon 
introduces singularities and there are no finite action solutions.   

As another example of a coordinate choice,
let $Gf(r)= -r/(1 - {\lambda \over 4}r^2)$ which gives the AdS space.  Then the curved BPS 
equations become:
 \be
h'(r) = -{(1 + {\lambda \over 4} r^2)\over r^2} (\varphi^2(r) -1)
\ee

\be
\varphi'(r) = -{1\over( 1 + {\lambda \over 4} r^2)} h(r) \varphi(r)
\ee
With this choice of coordinate, the line element becomes 
\be
(d s)^2 = {1\over{(1+ {\lambda \over 4}r^2})^2}\Bigg\{ r^2 d\Omega_2 +
 (d r)^2 \Bigg\}
\ee

\begin{figure}
\begin{center}  
\leavevmode
\epsfsize= 8.0in \epsfbox{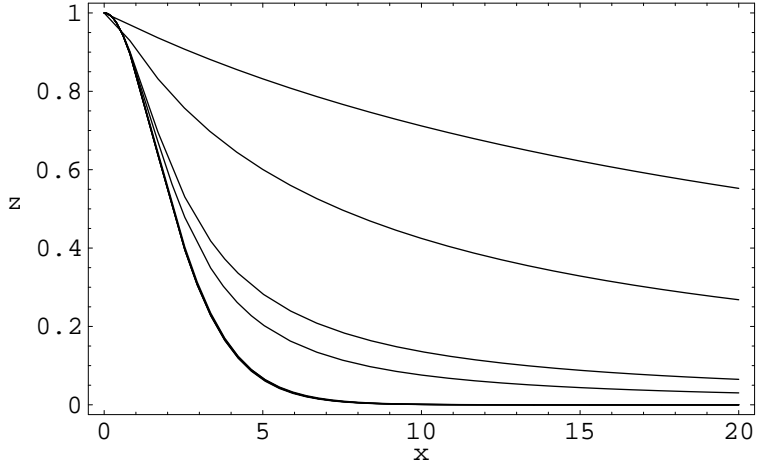}
\end{center}
\caption{ Non-zero component of the gauge field  $z(x)$ is shown for
various values of $\lambda$ 
for the choice $Gf(r)= -r/(1 - {\lambda \over 4}r^2)$.
The above
values, starting from the bottom correspond to $\lambda= 0, 4. 10^{-3}$
,$\lambda=0.1$, $\lambda=0.5$, $\lambda= 0.8$, $\lambda= 1$ and
$\lambda=10$ in units of $h_0^2$.}
\label{fig3}
\end{figure}

\begin{figure}
\begin{center}
\leavevmode
\epsfsize= 8.0in \epsfbox{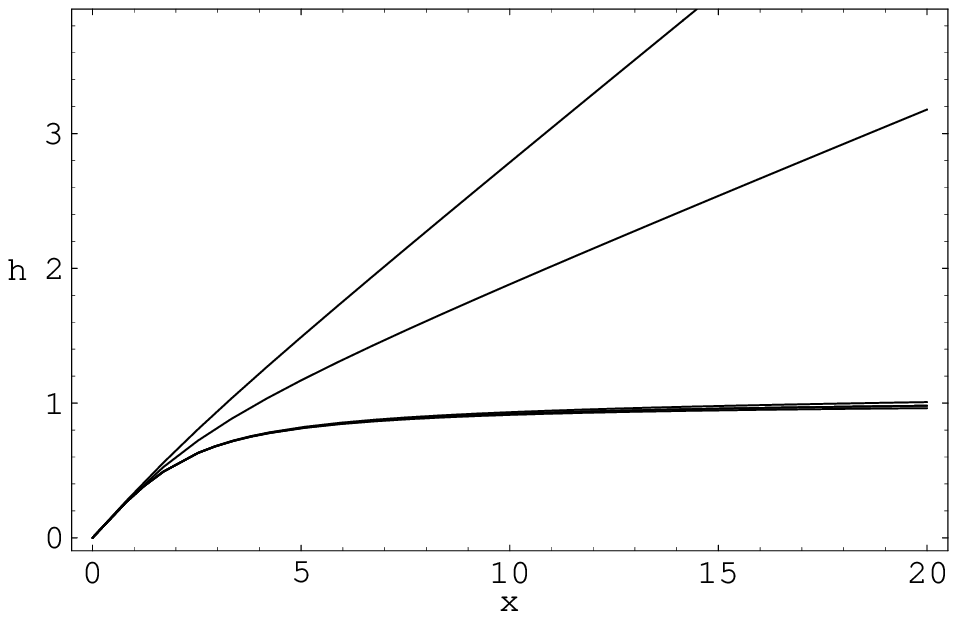}
\end{center}
\caption{ Higgs field $h(x)$ is shown for various values of $\lambda$
in the choice $Gf(r)= -r/(1 - {\lambda \over 4}r^2)$.
$h(\infty)$ approaches grows unbounded  if $\lambda$ is
increased. The above
values, starting from the top correspond to $\lambda= 0$, $4. 10^{-3}$
,$\lambda=0.1$, $\lambda=0.5$, $\lambda= 0.8$, $\lambda= 1$ and
$\lambda=10$  in units of $h_0^2$.}
\label{fig4}
\end{figure}

The numerical solutions to these equations are shown in \ref{fig3}
and \ref{fig4}. The gauge field behaves more or like the previous case but it approaches 
rather slowly to the vacuum solution, $( h(r) = 0, \varphi(r) = 1)$, when the 
cosmological constant is increased.
The Higgs field does not approach to the vacuum solution and it diverges when the cosmological 
constant is increased.

\section{Singular Solutions}
In this section we would like to point out an exact solution to equations 
of motion . 
These solutions are singular and have no
non-trivial limits in the flat space.
\be
\varphi(r)= -\sqrt{\lambda} G f(r), \hskip 1 cm h(r)= {{1 +\lambda G^2
f(r)^2}\over {f(r) G}}
\label{singular}
\ee 
For any choice of $f(r)$ it is not possible to meet the finite action
conditions for the gauge theory instantons. So these solutions are singular.

For definiteness let us rewrite these solutions
in the both coordinate examples chosen above.
For $ Gf = -r $ we have
\be
\varphi(r)= \sqrt{\lambda} r, 
\hskip 1 cm h(r)= -( \frac{1}{r} + \lambda r)
\label{singular3}
\ee

For the choice $Gf(r)= -r/(1 - {\lambda \over 4}r^2)$
\be
\varphi(r)= \sqrt{\lambda} {r \over (1 - {\lambda \over 4} r^2)}, 
\hskip 1 cm h(r)= -{(1 + {\lambda \over 4} r^2)^2 \over 
r (1 - {\lambda \over 4} r^2) }
\label{singular2}
\ee

These solutions are not gauge copies of the trivial vacuum solutions

\section{Conclusion}

We have shown that when three dimensional Euclidean gravity is
coupled to Yang-Mills and Higgs fields the equations of motion reduce
to first order equations of the Bogomol'nyi type. We found singular
and regular solutions. Our main result is that, if the three dimensional
space is a gravitational instanton, there are finite
action solutions for any positive semi-definite value of the cosmological
constant. Depending on the numerical value of the cosmological constant 
these solutions take values between the BPS solution and the trivial
vacuum solution. The action can be calculated exactly and is given by
(\ref{exactaction}) and the gravitational instanton action is (\ref{exactaction2}).
 Finite actions solutions are stable and
one can define a topological charge  which is  the magnetic
charge. This can be done following 't Hooft's definition of an Abelian 
field strength outside the instanton core.  

We have also showed that there if the cosmological constant is small ($ \lambda \le 0.8 h_0^2 $
 see fig 3 and 4) there are finite action solutions in the AdS space.
 The space itself has zero action.

In addition to the Yang-Mills term, if a Chern-Simons term is added for the
gauge sector one should look at the complex gauge configurations
as it was pointed out in \cite{Kamran}. 
In this case one cannot use a 
singular gauge as the Chern-Simons term trivially vanishes.
The theory will have the following additional action
\be 
S= -{i\kappa\over e^2}\int d^3 x \epsilon^{\mu \nu \lambda}
{\mbox{tr}}\bigg( 
A_{\mu} \dd_\nu A_\lambda + {2\over 3} A_\mu A_\nu A_\lambda \bigg)
\ee
Using the symmetric ansatz one gets
\be
S= {4i\pi\kappa\over e^2}\int_0^\infty\bigg[\epsilon_{\alpha \beta} 
\varphi_{\alpha}' \varphi_{\beta} +D (\varphi_{\alpha} \varphi_{\alpha} -1)
  - \varphi_2' \bigg]
\ee     
Clearly this term vanishes for the singular gauge which is too
restrictive. In some other gauges (i.e $D = 0$ and $\varphi_2 \ne 0$) we
expect complex solutions. For the case of Einstein-Maxwell-Chern-Simons
theory with a Lorentzian metric  we refer the reader to \cite{Fernando,Dereli}.

In the context of a non-supersymmetric theory (like the one we dealt with
in this paper) $h_0$ and $\lambda$ are given to define the theory. So one
changes the theory if these parameters are changed at the classical level.
So our results mainly mean that there are finite action instanton
solutions for those theories which have suitable pairs of $\lambda$ and
$h_0$. On the other hand if our theory is considered as a bosonic part of
a  Supergravity theory where a moduli space (for the Higgs field)  exists
then  for given $\lambda$ there are many solutions.

\section{Acknowledgements}
The research of B.\ T.\ is supported by  PPARC Grant PPA/G/O/1998/00567.
  The work of A.\ F.\ was 
supported in part by DOE grant DE-FG02-94ER-40823.

\vskip 1cm

\leftline{\bf References}  

\renewenvironment{thebibliography}[1]
        {\begin{list}{[$\,$\arabic{enumi}$\,$]}  % {\arabic{enumi}.}
        {\usecounter{enumi}\setlength{\parsep}{0pt}
         \setlength{\itemsep}{0pt}  \renewcommand{\baselinestretch}{1.2}
         \settowidth
        {\labelwidth}{#1 ~ ~}\sloppy}}{\end{list}}

\myend